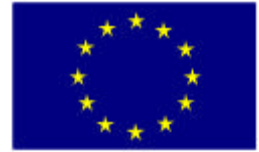

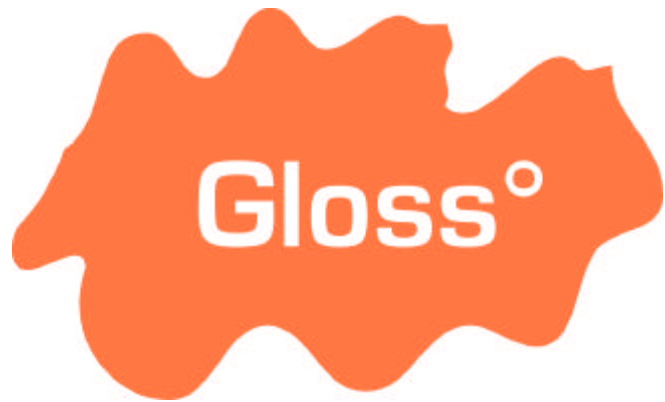

**Global Smart Spaces**

# First Smart Spaces

## D8, WP5

# 1 September 2002

GRAHAM KIRBY, ALAN DEARLE, ANDREW MCCARTHY, RON MORRISON, KEVIN MULLEN, YANYAN YANG, RICHARD CONNOR, PAULA WELEN, AND ANDY WILSON

| IST Project Number | IST-2000-26070 | Acronym | GLOSS |
|---|---|---|---|
| Full title | Global Smart Spaces | | |
| EU Project officer | Jakub Wejchert | | |

| Deliverable | Number | D 8 | Name | First Smart Spaces | | |
|---|---|---|---|---|---|---|
| Task | Number | T | Name | | | |
| Work Package | Number | WP | Name | WP5 Physical infrastructure | | |
| Date of delivery | Contractual | Project Month 21 | Actual | Project Month 21 | | |
| Code name | | | Version 1.0 | draft ☐ | final ☑ | |
| Nature | Prototype ☐ Report ☑ Specification ☐ Tool ☐ Other: | | | | | |
| Distribution Type | Public ☑ Restricted ☐ to: <partners> | | | | | |
| Authors (Partner) | University of St Andrews, University of Strathclyde, | | | | | |
| Contact Person | Al Dearle | | | | | |
| | Email | Al@dcs.st-andrews.ac.uk | Phone | | Fax | |
| Abstract (for dissemination) | This document describes the Gloss software currently implemented. The description of the Gloss demonstrator for multi-surface interaction can be found in D17. The ongoing integration activity for the work described in D17 and D8 constitutes our development of infrastructure for a first smart space. In this report, the focus is on infrastructure to support the implementation of location aware services. A **local architecture** provides a framework for constructing Gloss applications, termed **assemblies**, that run on individual physical nodes. A **global architecture** defines an overlay network for linking individual assemblies. Both local and global architectures are under active development. | | | | | |
| Keywords | Global architecture, local architecture, assemblies. | | | | | |

# 1 INTRODUCTION

This document describes the Gloss software currently implemented. The description of the Gloss demonstrator for multi-surface interaction can be found in D17. The ongoing integration activity for the work described in D17 and D8 constitutes our development of infrastructure for a first smart space.

In this report, the focus is on infrastructure to support the implementation of location aware services. A **local architecture** provides a framework for constructing Gloss applications, termed **assemblies**, that run on individual physical nodes. A **global architecture** defines an overlay network for linking individual assemblies. Both local and global architectures are under active development.

The structure of the software is outlined below. At the top level, the *infrastructure* package contains support for both local and global architectures; *model* contains the Gloss ontology; *services* contains the implementation of various web services; and *simulation* contains code for simulating various global architectural policies.

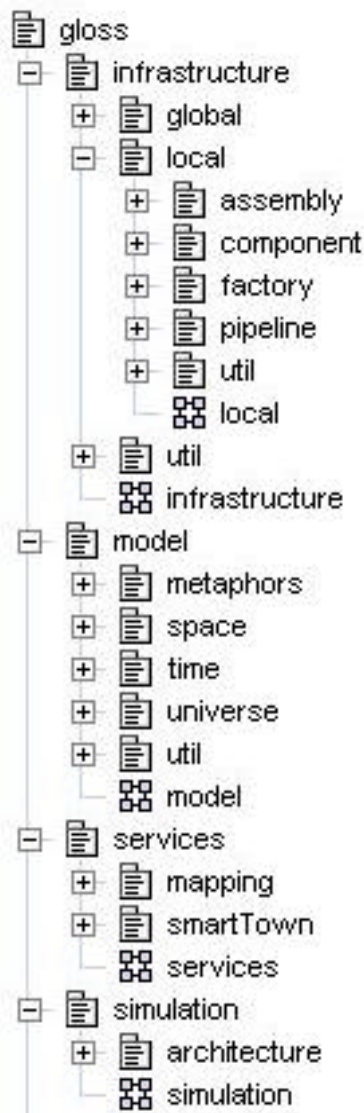

# 2 LOCAL ARCHITECTURE

Any particular location-aware service, such as *Radar* or *Hearsay* [1], will be implemented as a distributed collection of communicating assemblies. The local architecture defines a pattern for implementing individual assemblies.

## 2.1 PIPELINES

The local architecture, implemented in Java, is based on a pipeline of modular components. Events flow between components as strings, XML fragments or structured objects, as appropriate. The modularity facilitates the construction of various assemblies containing common components.

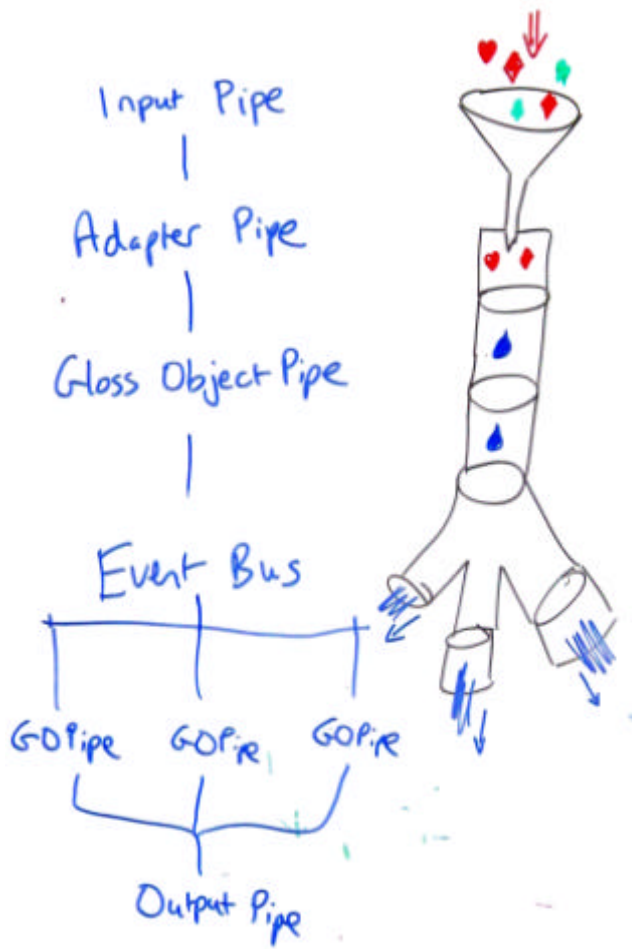

*Pipeline architecture for local assemblies*

The pipeline structure is defined by a number of interfaces. **Plug** interfaces allow components to be connected to upstream components. **PlugSocket** interfaces allow components to accept registrations from downstream components. There are variants for different types of events, currently either **String**s or **Object**s.

```
/** Interface defining pipeline components that can accept object
```

```java
 *  events from up-stream components. */
public interface ObjectPlug {

  /** Injects an object event into the component; invoked by the
   *  upstream component with which this component is registered. */
  void put(Object object);
}

/** Interface defining pipeline components that can accept
 *  registrations from down-stream ObjectPlug components. */
public interface ObjectPlugSocket {

  /** Registers a downstream ObjectPlug component in the pipeline.*/
  public void register(ObjectPlug objectPlug);
}

/** Interface defining pipeline components that can accept string
 *  events from upstream components. */
public interface StringPlug {

  /** Injects a string event into the component; invoked by the
   *  upstream component with which this component is registered. */
  void put(String string);
}

/** Interface defining pipeline components that can accept
 *  registrations from downstream StringPlug components. */
public interface StringPlugSocket {

  /** Registers a downstream StringPlug component in the pipeline.*/
  void register(StringPlug stringPlug);
}
```

Plugs and sockets are connected into pipelines as illustrated below:

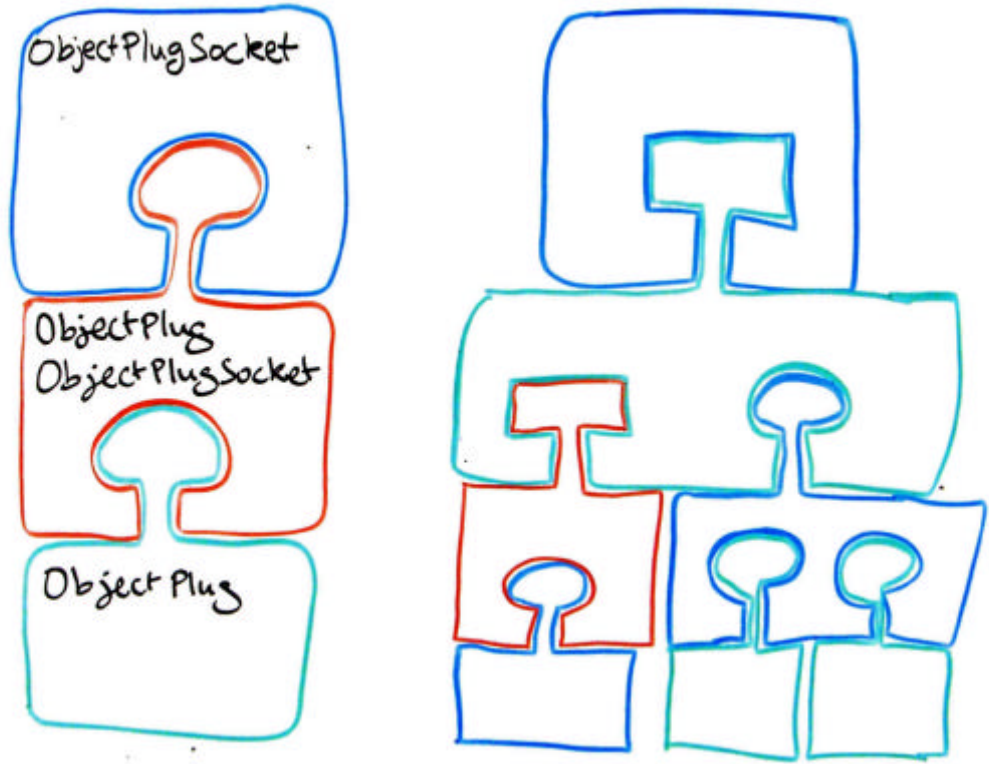

*Components in pipelines*

In the simple pipeline on the left, object events are passed progressively down the linear pipeline. The more complex example on the right shows how the event types may be transformed by adapters as the events flow down the pipeline, and how there may be forks in the pipeline due to components that implement multiple interfaces.

### 2.2 PIPELINE COMPONENTS

Components currently implemented provide GPS and SMS wrappers, XML storage and retrieval, and an 'event bus' to allow delivery of events to multiple consumers.

```
/** Pipeline component implementing an event bus, with which
 *  multiple downstream components may register. Events received
 *  from upstream are distributed to all registered components. */
public class EventBus implements ObjectPlugSocket, ObjectPlug ...

/** Pipeline component representing a GPS device; emits
 *  LatLongCoordinate events at user-determined intervals. */
public class GPSDevice implements ObjectPlugSocket ...

/** Pipeline component representing an SMS-capable device; accepts
 *  and emits arbitrary string messages. Messages flowing into the
 *  component are sent via SMS to an SMS gateway. SMS messages
 *  received by the component are injected into the pipeline. */
public class SMSDevice implements StringPlugSocket, StringPlug ...
```

```
/** Pipeline component representing an SMS-capable device; accepts
 *  and emits String messages, which must be valid XML fragments.
 *  Messages flowing into the component are sent via SMS to an SMS
 *  gateway. SMS messages received by the component are injected
 *  into the pipeline. */
public class SMSXMLDevice extends SMSDevice
    implements StringPlugSocket, StringPlug ...

/** Pipeline component that records each string received from
 *  upstream into a date-stamped file. */
public class SaveStringToDateStampedFile implements StringPlug ...
```

## 2.3 ASSEMBLIES

With appropriate reuse of components, simple assemblies can be defined using only a few lines of code; assemblies currently in use include a mobile assembly running on a Pocket PC to transmit GPS-derived location information via SMS, and a server-based assembly for processing and storing such location events. These are motivated by the use case outlined below:

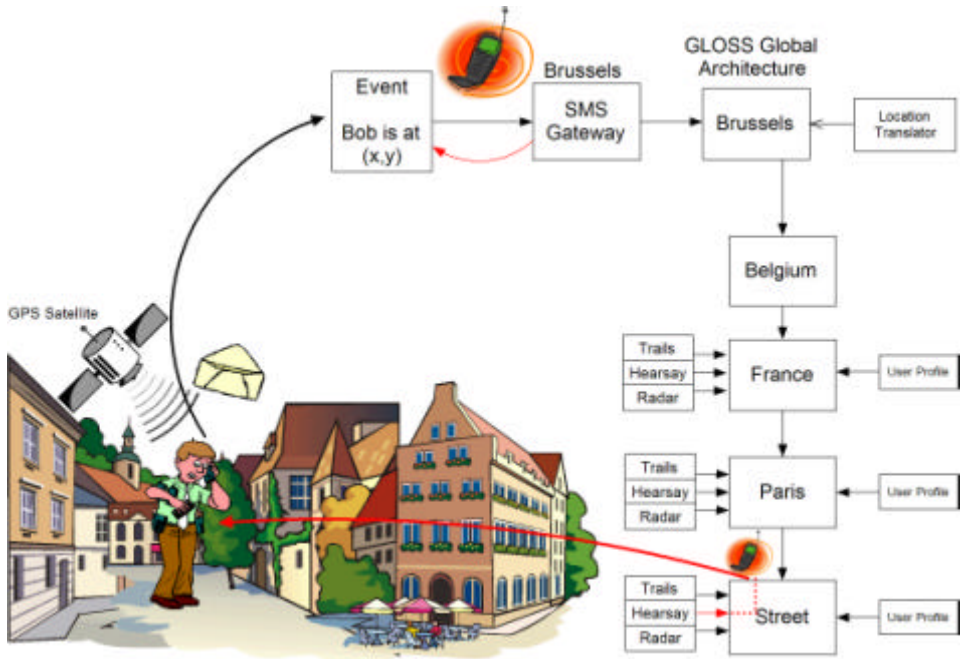

*A motivating use case. Bob transmits his location to the GLOSS infrastructure via his mobile device. This enables various services: relevant hearsay is delivered to him when he enters the corresponding proximity; his trail is accumulated and may be observed by others (with appropriate permission); his radar indicates the relative locations of various landmarks, commercial services and other mobile users.*

The figure below shows a user interface tool for dynamically creating assemblies. In many Gloss applications we envisage assemblies being created by an assembly program (an application factory).

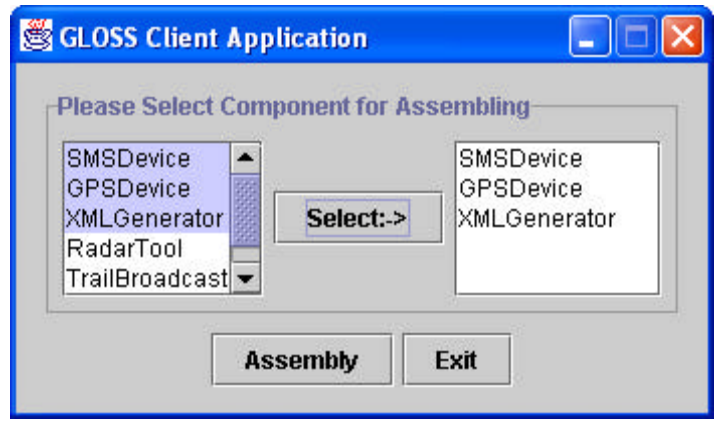

*Prototype user interface for dynamic construction of new assemblies*

### 2.3.1 MOBILE ASSEMBLY

The mobile assembly currently in use generates XML-encoded user location events, for use by various location-aware services such as trails, radar and hearsay. The events are transmitted via SMS to a location database server.

The physical components are a Compaq iPaq handheld PocketPC, a compact flash format GPS device, and a PCMCIA format GSM device. The assembly is shown below. The various adapters and extenders are needed to overcome a problem caused by the close physical proximity of the two PCMCIA slots on the PocketPC jacket.

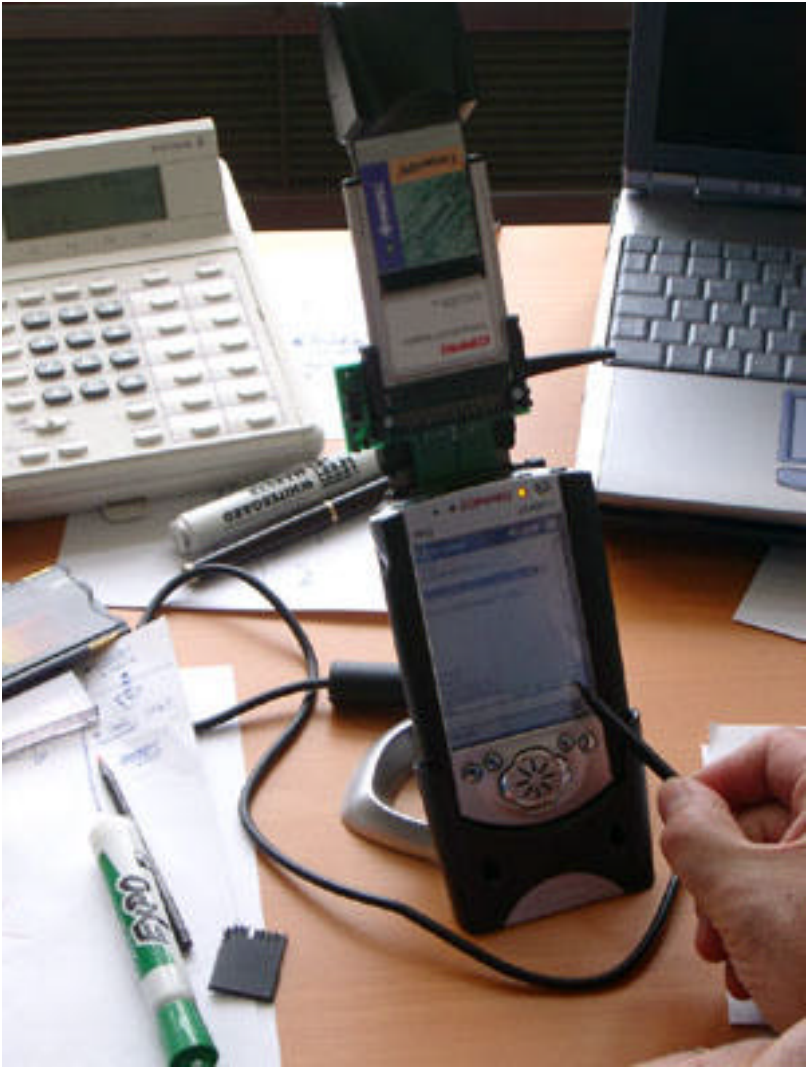

*Close-up of mobile assembly*

In the next version, the handheld will communicate with a separate GSM phone via BlueTooth, to overcome the physical awkwardness of the current prototype.

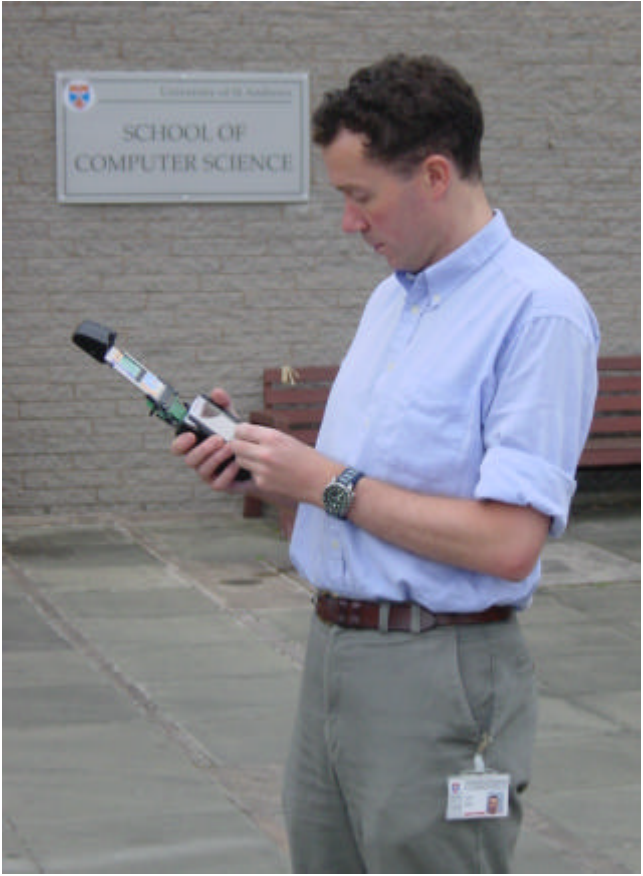

*Mobile assembly in use*

The mobile assembly is programmed in Java using IBM's WebSphere Studio Workbench environment [3]. The constituent pipeline components are:

- a wrapper for the GPS device, emitting Latitude/Longitude location objects;
- an XML generator that converts location objects to XML strings;
- an event bus;
- a wrapper for the GSM/SMS device, that sends string messages to a remote server;
- adapters that convert between objects and strings.

The Java code to create and run the mobile assembly is as follows:

```java
/** Create the components. */
GPSDevice    GPS_device    = new GPSDevice();
XMLGenerator XML_generator = new XMLGenerator();
GPSAdapter   GPS_adapter   = new GPSAdapter();
EventBus     event_bus     = new EventBus();
SMSAdapter   SMS_adapter   = new SMSAdapter();

StringPlug SMS_device =
  SMSDeviceFactory.createSMSDevice(GSMSerialConnection.POCKET_PC);

/** Assemble the pipeline. */
GPS_device.register(XML_generator);
XML_generator.register(GPS_adapter);
```

```
GPS_adapter.register(event_bus);
event_bus.register(SMS_adapter);
SMS_adapter.register(SMS_device);

/** Start the assembly. */
GPS_device.run();
```

In this example the event bus is not strictly necessary since only one downstream component is registered with it, but is retained for generality.

### 2.3.2 SERVER ASSEMBLIES

The server assembly corresponding to the mobile assembly described above runs on a PC equipped with a GSM card. It receives incoming SMS messages and stores the user location information as date-stamped XML documents in a file directory. The code to create and run the server assembly is as follows:

```
/** Create the components. */
SMSDevice SMS_device =
  SMSDeviceFactory.createSMSXMLDevice(GSMSerialConnection.WIN32);

StringPlug saviour = new SaveStringToDateStampedFile();

/** Assemble the pipeline. */
SMS_device.register(saviour);

/** Start the assembly. */
SMS_device.start();
```

The XML file repository is accessed by a web service running on the same machine as the server assembly, or another that can mount the same file system. This service monitors the file system and automatically loads any incoming XML files, first as DOM objects and then, via a collection of GLOSS object factories, into instances of classes in the ontology [4] used to model the GLOSS domain. The graph of loaded objects, representing the server's current knowledge of the world, may be queried via a web form. Experiments with various services that may be provided on top of this are ongoing.

The queries currently supported are:

- **user location**: this retrieves the most recent known location for a particular user, specified using a GSM phone number as an approximation to a Globally Unique Identifier (GUID), and if possible displays the position on a map. Maps may be obtained dynamically from various public online services, or specified as static maps stored on the server.

- **user trail**: this renders the sequence of all known locations for a particular user as a trail over a map of suitable size.

- **smart town**: this indexes a user's current location against a database containing information on local facilities and commercial services, and returns a linked set of web pages that enable this to be traversed.

The example below shows the result of a user location query:

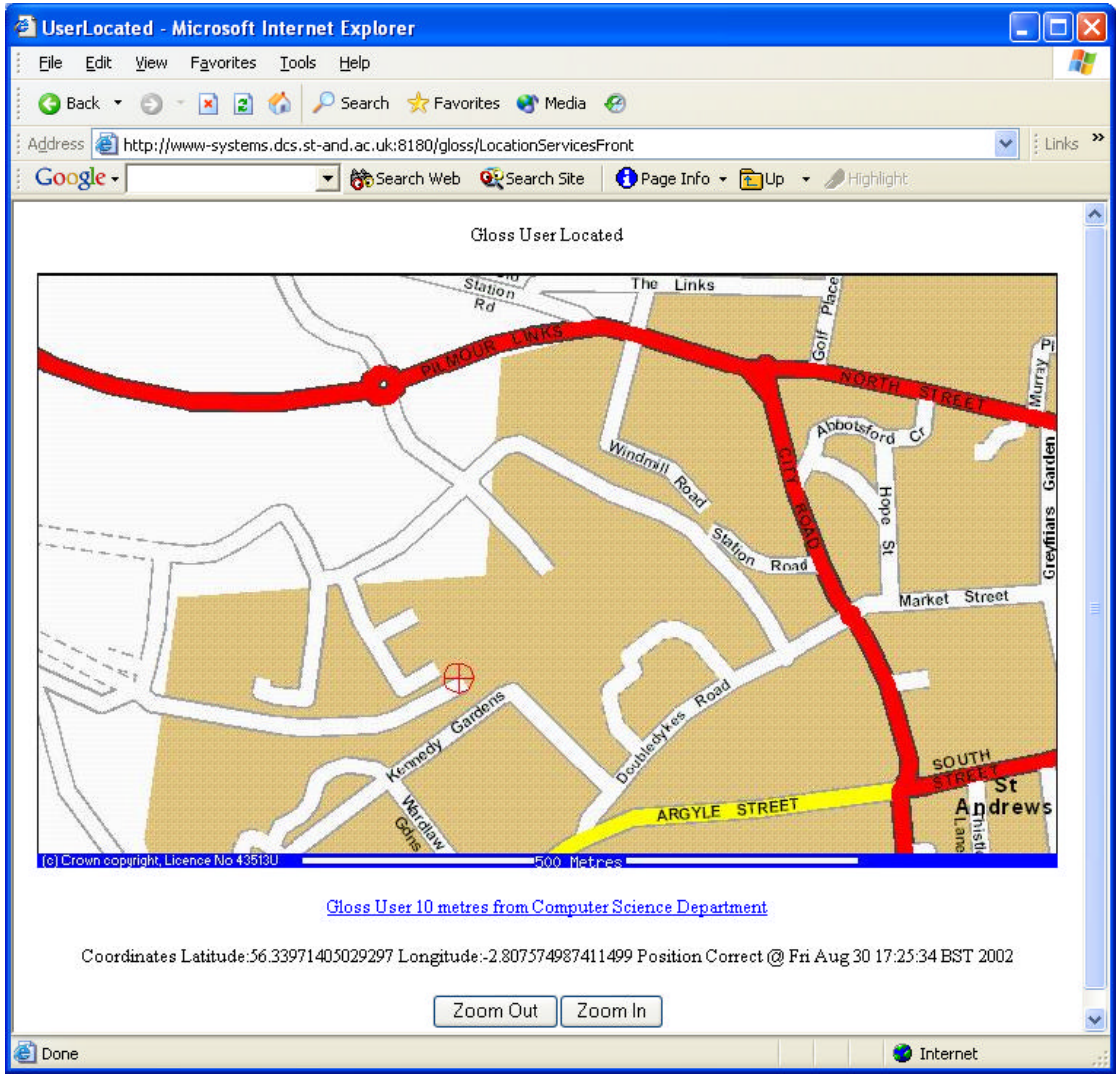

*User location web service*

The following example below shows the result of a smart town query:

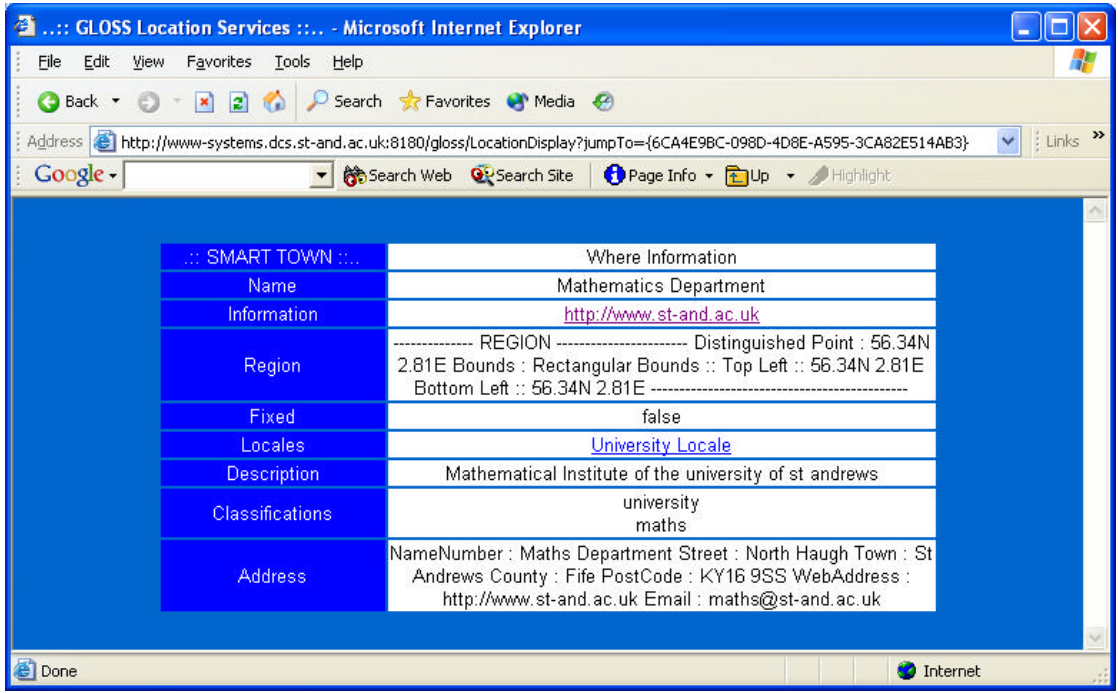

*Smart town web service returning Information relevant to user's current location*

# 3   GLOBAL ARCHITECTURE

Global location-aware services will require deployment of large number of geographically disparate assemblies, with events flowing both within and between assemblies. We do not yet take a strong position on the structure of such a global network, indeed it seems likely that optimal configurations may vary markedly for different services. Currently we assume only the following:

- GLOSS will require an overlay network of assemblies communicating over various transport mechanisms such as IP, SMS, Bluetooth, passive proximity detectors etc.
- The network will be heavily decentralised, perhaps exploiting current activity in peer-to-peer routing algorithms.

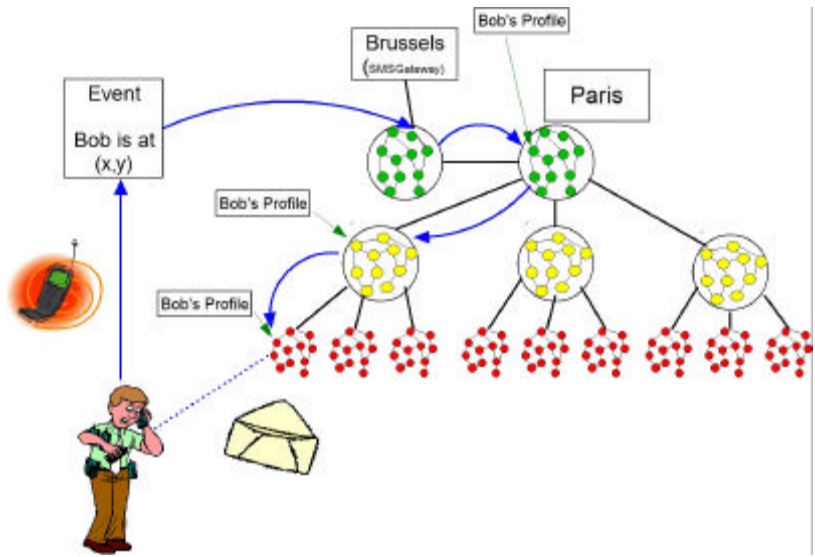

*A possible global architecture supporting use case*

To investigate various strategies for global deployment, we have developed a simulation infrastructure that allows us to describe and run particular services in terms of a topology of assemblies, the event message types and the local processing triggered by each event. Such a network can then be simulated on a single physical node, or deployed across a Beowulf cluster. The example below shows the output from the simulator:

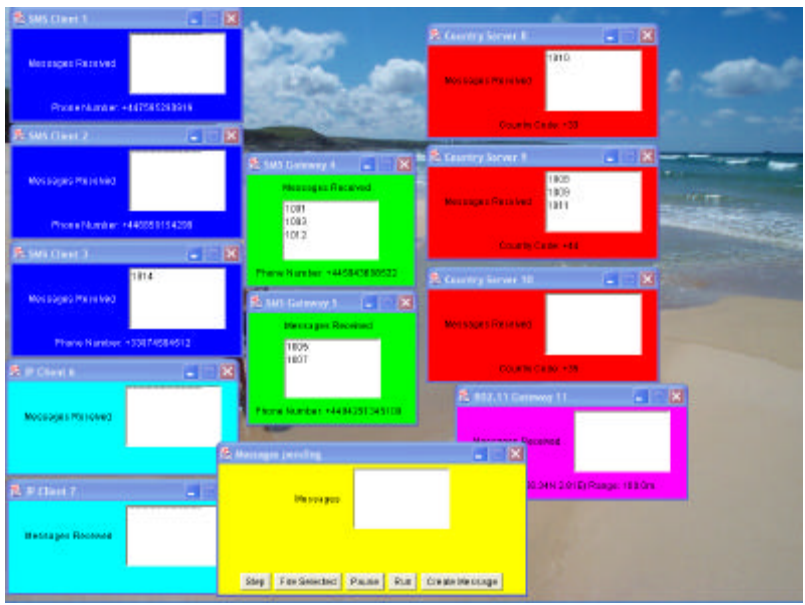

*Simulation of a particular global architecture*

## 4   REFERENCES

[1] A Recursive Software Architecture for Location-Aware Services. Alan Dearle, Graham Kirby, Ron Morrison, Kevin Mullen, Yanyan Yang, Richard Connor, Paula Welen, and Andy Wilson. GLOSS internal report.